# Active Physics-Informed Deep Learning: Surrogate Modeling for Non-Planar Wavefront Excitation of Topological Nanophotonic Devices

Fatemeh Davoodi[1,2,]*
[1]Institute of Experimental and Applied Physics, Kiel University, 24098 Kiel, Germany
[2]Kiel Nano, Surface and Interface Science KiNSIS, Christian Albrechts University, Kiel, Germany

Email: davoodi@physik.uni-kiel.de

**Abstract**
Topological plasmonics offers new ways to manipulate light by combining concepts from topology and plasmonics, similar to topological edge states in photonics. However, designing such topological states remains challenging due to the complexity of the high-dimensional design space. We present a novel method that uses supervised, physics-informed deep learning and surrogate modeling to design topological devices for specific wavelengths. By embedding physical constraints in the neural network's training, our model efficiently explores the design space, significantly reducing simulation time. Additionally, we use non-planar wavefront excitations via electron beams to probe topologically protected plasmonic modes, making the design and training process nonlinear. Using this approach, we design a topological device with unidirectional edge modes in a ring resonator at specific operational frequencies. Our method reduces computational cost and time while maintaining high accuracy, highlighting the potential of combining machine learning and advanced techniques for photonic device innovation.

In recent years, topological plasmonics has emerged as a transformative field in the manipulation of light on the nanoscale.[1] Topological systems, known for their robust nontrivial characteristics that remain unaffected by perturbations, have attracted significant attention across multiple scientific fields.[2–5] These systems exhibit remarkable phenomena, such as protected edge states, making them highly promising for applications in nanophotonics, quantum information processing, and other cutting-edge technologies. One of the pioneering models in topological plasmonics is based on the Su-Schrieffer-Heeger (SSH) model[6–9] often implemented using metallic nanoparticle chains with alternating spacings.[7,10–13] However, the fabrication precision required for nanoparticle interactions poses challenges, prompting the use of perforated nanoholes instead. Yet, relating the resonance frequencies of nanodisks with their hole counterparts and finding the topological modes of nano-hole chains at desired frequencies is very complex due to the differently the near-field interaction of local electromagnetic fields nanodisks and nanoholes.[14–16] Achieving the advantages of topological plasmonic chains, particularly in mesoscopic SSH rings, requires symmetry-breaking excitation techniques using non-planar wavefronts. On the other hand, designing topological photonic systems with non-planar wavefront excitations adds another layer of complexity , as the interaction between these wavefronts and the system's topological and trivial modes must be carefully managed.[7,17,18] This complexity is amplified when attempting to excite modes at specific wavelengths, where non-linear interactions demand advanced modeling and simulation techniques.[19–23] Optimizing geometries based on the traditional methods for specific photonic functions involves thousands of iterations, posing significant challenges.[18,24] In this context, incorporating deep learning methodologies offers a transformative solution.[25–27] Physics-informed surrogate models provide an efficient approach for predicting device characteristics and generating optimized topologies for multiple objectives.[28–32]

One of the key advantages of surrogate models as a data-driven approximation of complex physical systems is their ability to capture non-linear relationships between device parameters (e.g., size, spacing, material properties) and optical performance.[33–35] By training on a limited set of simulation data, they enable efficient exploration of the parameter space, making it easier to discover optimal designs for specific applications like sensors, waveguides, or optical switches.[36] However, traditional surrogate models often require large datasets and lack the integration of underlying physical knowledge, limiting their effectiveness.[37–39] To overcome this, physics-informed machine learning (PIML) has emerged as an approach that combines physical laws with data-driven models, improving the ability to generalize even with smaller datasets.[40,41] By incorporating physical constraints like plasmonic mode coupling and resonant behavior, our model can rapidly approximate topological properties based on input parameters such as nanoparticle size, geometry, and coupling strength. However, the system exhibits highly nonlinear behavior due to three primary factors: strong coupling of nano resonators, topological plasmonic modes, and non-uniform excitation. These factors result in a significant computational load, with simulations for just one structure taking several hours or even days on powerful servers, making large-scale simulations resource-intensive and costly. Additionally, to avoid overfitting in such complex predictions, a sufficiently large dataset is needed. However, generating this dataset through thousands of iterations is nearly impractical. Therefore, we choose an optimized method to predict the desired properties more efficiently. We trained our model in three stages using a relatively small dataset of 4,000 simplified simulations per phase. Guided by established physical principles, the model accurately predicts optimal designs that achieve robust one-way edge modes at specific wavelengths.

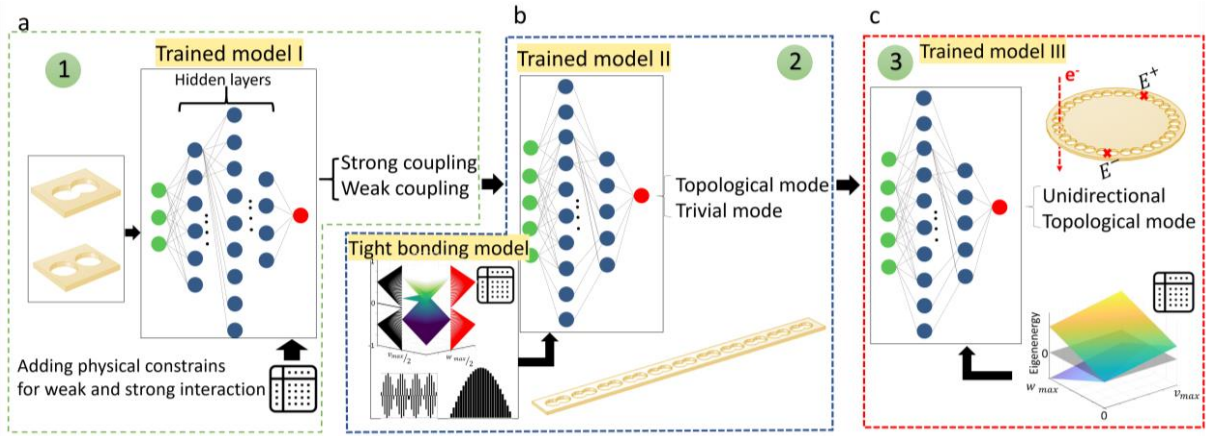

Figure 1: The overview of the three-step model training process for predicting topological modes in physical systems. (a) Phase I: The model is trained on strong and weak coupling, introducing physical constraints for interactions between unit cells. (b) Phase II: The model learns topological phases and their associated modes using a tight-binding model to distinguish between topological and trivial modes. (c) Phase III: The model identifies topological and trivial directions in a ring, with unidirectional topological mode propagation under non-uniform excitation.

This physics-informed approach ensures that predictions remain consistent with fundamental physical laws, enabling fast and accurate design of topological photonic devices, even without extensive simulation data. In our device design process, the surrogate model replaces full-scale COMSOL simulations by approximating structural properties using input parameters such as nanoparticle size, coupling strength, and geometry. At the initial stage, we generate a dataset of 4000 simulations to train the model on strong and weak coupling scenarios. We use a simulation-driven approach to train a surrogate model for predicting plasmonic behavior in nanohole arrays, focusing on extinction cross-sections $\sigma_{ext}(\lambda)$ as a function of the hole diameter ($D$), the inter-hole distance ($G$), and the wavelength $\lambda$, the structure here excited with plane wave in the direction depicted in Figure 2. The simulation is performed using COMSOL and generates datasets that enforce physical constraints in both strong and weak coupling regimes. By integrating a prepared table with labeled data during the training phase, the surrogate model ensures that predicted interaction strengths align with expected behaviors in both strong and weak coupling regimes.

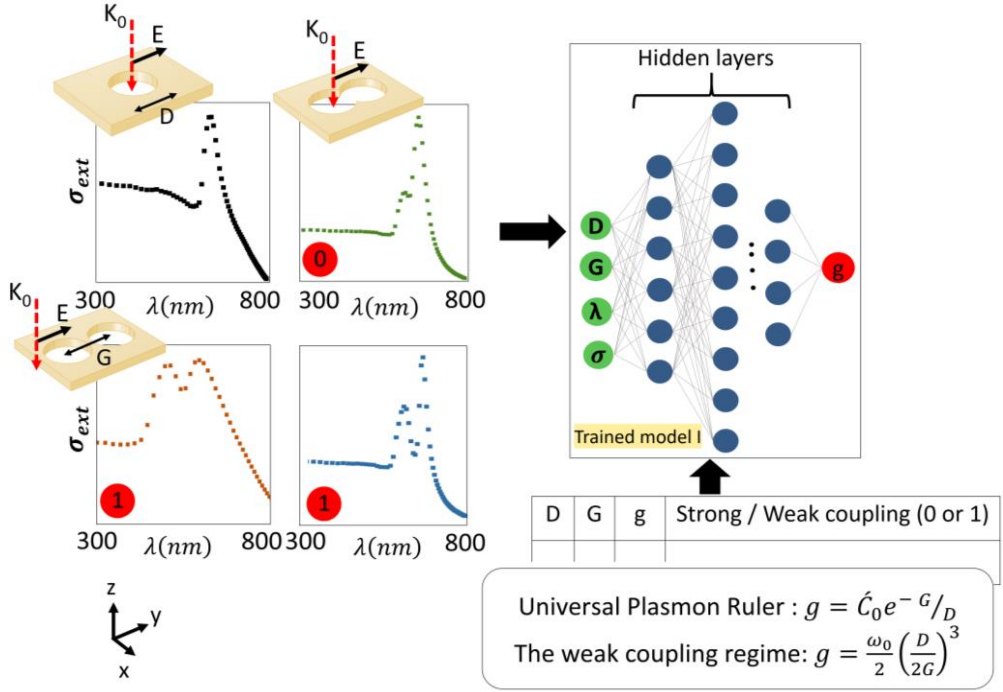

Figure 2: The extinction spectra with plane wave excitation for (a) a single nano resonator, (b) symmetric merged nano resonator dimer and (c, d) a dimer of gold nanoparticles with small gap distances. The inset labeled with 0 and 1 for strong and weak coupling regimes. (e). A neural network with 4 inputs, 5 hidden layers and one output which is trained to distinguish between strong and weak coupling regimes, using labeled data (0 for weak coupling and 1 for strong coupling). This setup introduces a supervised learning framework, allowing the model to classify these physical states based on the nonlinear behaviors seen in the extinction spectra.

This technique acts as a regularizer, reducing the need for computationally expensive full simulations by enabling rapid predictions of new parameter sets. It guarantees the network produces physically consistent outputs, particularly with regard to the bonding and antibonding plasmonic modes in strong coupling. Compared to purely data-driven models, this approach is more effective in maintaining physical accuracy, as it directly enforces the expected interaction patterns. This approach is particularly useful in fields like nanophotonics, where traditional simulation methods are resource-intensive. The input dataset for strong coupling interactions is established using the Universal Plasmon Ruler Equation, which relates the shift in the plasmon resonance wavelength ($\Delta\lambda/\lambda_0$) to the inter-particle distance ($G$) as follows: $\frac{\Delta\lambda}{\lambda_0} = C_0 e^{-G/D}$, $C_0$ is scaling constant.[42] In contrast, for weak coupling, the interaction strength g between two nanoparticles follows an inverse cubic dependence on the distance $G$, expressed as $g = $ $\frac{\omega_0}{2}\left(\frac{D}{2G}\right)^3$.[15,43] We have classified the coupling regime (strong or weak) in our dataset based on the corresponding extinction cross-section data, when the splitting in the resonance peaks appear, Label: 1 ("strong"), in the case a slight shift in the resonance wavelength (no distinct mode splitting), Label: 0 ("weak"). This gives us a direct mapping from the distance $G$ between nanoholes to the interaction strength $g$. In this step, the model learns both: (i). the transition point from weak to strong coupling as the distance changes. (ii). the quantitative relationship between the distance $G$ and coupling strength $g$ using both the weak coupling formula and the Plasmon Ruler for strong coupling.

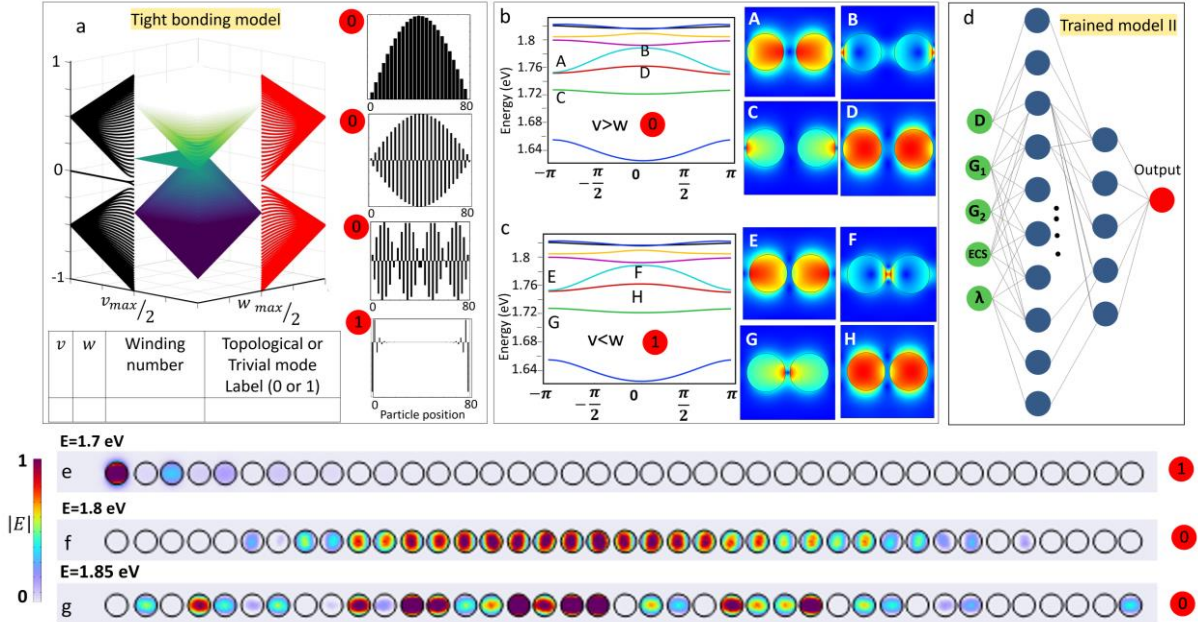

Figure 3: (a). The energy spectra of the SSH Hamiltonian for a truncated lattice comprising 80 unit cells versus the coupling constant between the nanoparticles in a plasmonic dimer calculated using a nearest-neighbor tight-binding approach. The horizontal axis is normalized relative to the maximum coupling constant. The energy eigenvectors of the structure calculated at the coupling strength corresponding to the bulk and the edge modes. The inset labeled with 0 and 1 for bulk and edge modes. (b, c). The numerically simulated band structures are computed for the trivial lattices ($v > w$) and the nontrivial ($v < w$) in the momentum space. (d). A neural network with 5 inputs, 3 hidden layers and one output which is trained to distinguish between topological and trivial modes in truncated SSH chain created with 40 sites perforated holes inside 50 nm thick gold layer (0 for trivial mode and 1 for topological mode). The spatial profile of the electric field is shown for the edge mode (f) and two bulk modes (e, g) at depicted energies for a plasmonic SSH ring chain structure comprising 40 unit cells. The parameters $d = 250$ nm, $h = 50$ nm, $G_1 = 260$ nm, and $G_2 = 240$ nm, respectively, represent the diameter, height, center to center distances of nano resonators which labeled with 0 and 1 for bulk and edge modes.

In the second phase of training, we expand the deep learning model's scope by incorporating simulation data from SSH chain configurations. The aim is to train the model to identify structural configurations that support topological modes under varying physical conditions, such as hole diameter, inter-hole spacing, extinction cross-section, and wavelength. This additional training enables the model to identify topological phases and associated modes, in conjunction with learning from tight-binding calculations[44] that involve hopping parameters and winding numbers. The winding number is employed as a topological invariant, helping to distinguish between topological and trivial modes,[7] as illustrated in Figure 3(a). A comprehensive dataset has been developed where each entry corresponds to a specific set of parameters $v$ and $w$, along with their respective winding number $W$, and topological phase for various chain sizes (for additional details, see Supplementary Sections I and II). The appearance of localized edge modes at the boundaries of the system is a direct signature of a non-trivial topological phase.

By adding tight bonding Table $T_1$, during the training phase of surrogate model, the model enforces that the predicted topological phase transitions the expected behavior based on the winding number. we labeled the data points in our dataset with the corresponding phase based on the relationship between $v$ and $w$ and the calculated Winding number: (i). Topological Phase: Occurs when $v<w$, meaning the coupling between adjacent nanoparticles in the chain is weaker than the coupling across neighboring dimers ($W>1$). (ii). Trivial Phase: Occurs when $v>w$, meaning the coupling between adjacent nanoparticles is stronger than the intra-dimer coupling ($W=0$).[13] The tight-binding model is in the quasistatic limit and dipole approximation,[31-32] where the nanoparticles are lossless and chain dimensions are much smaller than the wavelength. Therefore, we used the infinite chain of perforated holes simulations dataset for limited numbers of $G_1$ and $G_2$ as the intercell and intracell distances between the holes. The photonic band structures, shown in Figure 3b, reveal both labeled topological and trivial photonic band gaps. At the last step of this step, we simulate a truncated SSH chain based on perforated holes inside a 50 nm gold layer with distances $G_1$ and $G_2$ in real-space simulations using COMSOL. We have trained the model to recognize edge state localization by using the localization length (intensity of the wavefunction at the boundary) (Supplementary Figure $S_2$). As shown in the Figure 3c, to train the model on this additional phase information, we follow a similar strategy to the previous coupling regime classification: our pretrained model has been trained to connect $v$, $w$, strong and weak coupling to topological/trivial phases. The model in this phase has five inputs, the diameter $D$ of the perforated hole in 50 nm gold layer, the distance $G_1$ and $G_2$ between two interaction holes, the wavelength $\lambda$, and the extinction cross section $\sigma_{ext}(\lambda)$. The output Label each data point as either topological (1) or trivial (0). By training the model on both coupling regimes and topological phases, the model can predict whether a configuration falls into the topological or trivial phase and whether it exhibits strong or weak coupling.

In the third and final phase of training, we extend the model to incorporate complex, non-planar excitations in a ring geometry of nano-hole chains based on the SSH model. This step emphasizes the design, characterization, and manipulation of topological plasmonic chains in scenarios involving non-planar wavefront excitations such as electron beams and quantum dots, which introduce a new layer of complexity. The inherent challenges associated with non-planar excitation arise from the nonlinear interactions between the excitation source (e.g., electron beam or quantum dot) and the nano-hole structure. This requires careful consideration of both the interaction geometry and the response of the topological system. In particular, the electron beam interacts with the structure in a nonlinear manner, necessitating sophisticated design strategies to maintain robust topological behavior under these unconventional conditions. For this phase of the deep learning model, we build upon the pretrained model from earlier stages, focusing on a dataset derived from simulations of SSH chains arranged in a ring geometry and non-planar excitations (see Supplementary Section III). Each unit cell in the SSH chain contains two nano-holes, positioned according to specific geometrical constraints. In this phase, the model is trained with a dataset containing variations in several key parameters: (i). Number of unit cells $N$: Different values for $N$ affect the localization and protection of topological modes. (ii). Radius of the ring $b$: This influences the interaction strength between unit cells and the geometry of the plasmonic modes. (iii). Hole diameter $D$: This parameter directly impacts the plasmonic resonance frequencies and the coupling behavior between the nano-holes. The extinction cross-section, wavelength, and the resonance condition of each hole are used as additional input features to the model. To further enhance topological design, we incorporate data generated from tight-binding calculations in a ring geometry based on the parameters $v$ and $w$, which correspond to the intra-cell and inter-cell coupling strengths, respectively. This allows the model to learn the relationship between the topological winding number and the geometry of the chain. In this phase, we impose additional physics-informed constraints on the model, focusing on the winding number as a key criterion for determining the topological nature of the designed SSH ring structures.

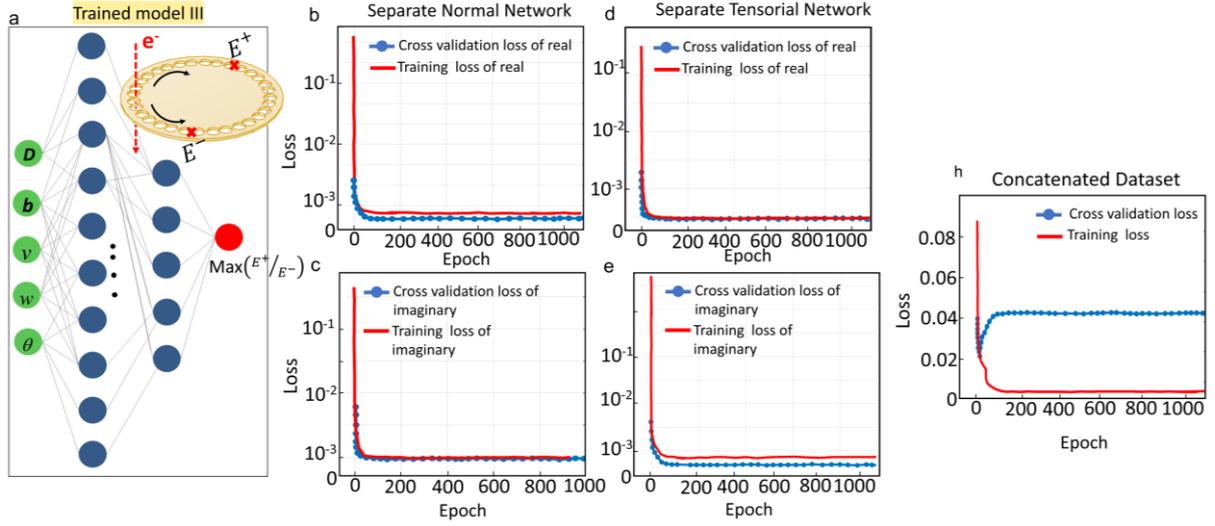

Figure 4: The third step of the model training. A neural network with 5 inputs, 3 hidden layers and one output which is trained to distinguish between topological and trivial modes in truncated SSH chain created with 40 sites perforated holes inside 50 nm thick gold layer (0 for trivial mode and 1 for topological mode). (a) Schematic of the neural network model (Trained model III) applied for analyzing the coupling between plasmons in a nanoring system. The model takes input parameters such as the big ring radius ($b$), the hole dimeter ($D$), the intercell and intracell interaction strength ($v$, $w$), and the dimerization parameter ($\theta$) to output the ratio Max($E^+/E^-$) for predicting uni-directional topological propagation. (b–e) The performance of separate neural network architectures for predicting the real and imaginary parts of the system's response. Subfigures (b, c) show the loss functions of the Separate Normal Network for the real and imaginary components, respectively, with both training and cross-validation loss converging smoothly. Subfigures (d, e) display similar results for the Separate Tensorial Network. (h) Loss performance for the concatenated dataset approach, showing slower convergence of the cross-validation loss compared to the individual networks, highlighting potential overfitting with 4000-dataset to the training data.

The winding number $M$ represents the number of times that the phase $\varphi$ winds around the circle as $W$ traverses a complete Brillouin zone.[45]

$$W = \frac{1}{2\pi} \int_{BZ} \frac{\partial \phi(k)}{\partial k} dk \tag{1}$$

This parameter directly correlates with the loop number $M$ in the SSH ring and is defined by the equation:[7]

$$E\big(\phi(k) + \delta\phi(k)\big) = e^{iM\delta\phi(k)} E\big(\phi(k)\big) \quad M = 0,1,\ldots,N-1 \tag{2}$$

where $\delta\phi(k) = \frac{2\pi}{2N}$ is the phase shift between two particles and $N$ is the number of unit cells.
In order to train the model for uni-directional propagation in a ring geometry, we label each position within the SSH ring model dataset. These labels link the propagation direction to the topological and trivial phases, as illustrated in Figure S5. The ring structure is excited by a nano-planar point excitation, an electron beam, at specific locations. To detect the topological one-way edge mode, we place two probes symmetrically on opposite sides of the ring for output. Let $E_1$ and $E_2$ represent the electric fields measured at these two points. The model outputs the ratio: $E^+/E^-$. If this ratio is maximized, it indicates that the excitation has successfully broken the symmetry and that the energy is predominantly propagating in one direction, confirming the existence of a topological mode.

This provides direct evidence of the structure supporting robust one-way edge propagation, a hallmark of topologically protected modes. To accurately predict optical properties and distinguish between topological and trivial modes, the model must handle complex output values. We explored three distinct approaches to structure the learning loss and optimize the DNN for complex-valued outputs: (i). Concatenated Method (CM): In this method, we treat the real and imaginary parts of the complex output as a single vector.[46] The real and imaginary components are concatenated into one feature, and the DNN learns to predict this combined feature. The concatenated approach simplifies the optimization process by reducing it to a single loss function, thereby enabling efficient learning of complex-valued properties. (ii). Tensorial Separated Method (TSM): In this approach, we explicitly separate the real and imaginary components and treat them as independent tensors. The DNN predicts the real and imaginary parts separately but within the same model.[47–49] This method ensures that each component receives individual attention during the optimization process, while still being interconnected during the learning process. (iii). Normal Separated Method (NSM): Here, the real and imaginary parts are treated as completely separate outputs. The DNN has two distinct output layers: one dedicated to predicting the real part and another for predicting the imaginary part. This method provides more flexibility in the model's architecture and allows for independent learning of both components, which is especially useful when the real and imaginary parts exhibit different behaviors.[48] The evaluated loss function are compared in Table I. The TSM method allowed the network to better capture the relationship between the real and imaginary parts of the system's response, leading to more accurate predictions. We employed the Adam optimizer to efficiently minimize the discrepancy between the predicted and labeled values of the model output.

Table I: The training and cross validation loss for NSM, TSM methods.

| parameter | Loss performance of NSM (training/cross validation) ($\times 10^{-3}$) | Loss performance of TSM (training/cross validation) ($\times 10^{-3}$) | Loss performance of CM (training/cross validation) ($\times 10^{-2}$) |
|---|---|---|---|
| Real | 1.93/1.29 | 0.14/0.133 | 1.67/4.13 |
| Imaginary | 1.96/1.32 | 0.141/0.13 | |

The DNN was trained to predict the ratio of the maximum electric field values for positive and negative, $E^+/E^-$, based on key parameters, the training process revealed optimal conditions for maximizing this ratio, occurring at a frequency of 380 THz, with a nanohole diameter $D$=248 nm and a ring diameter $b$=1380 nm. These findings are presented in Figure 5 (a-c), showcasing the peak response across the parameter space. To confirm the accuracy of the model, we conducted a full-wave simulation at the identified optimal conditions. The simulation, illustrated in Figure 5 (d-f), demonstrated the unidirectional propagation of a topologically protected mode with dipole excitation at this frequency. To assess the robustness of this mode, we introduced two types of defects: one by reducing the size of a nanohole and another by completely omitting one nanohole. Despite these defects, the mode continued to propagate without significant disruption, a key indicator of topological protection. Furthermore, we calculated the vorticity of the spin angular momentum (Figure 5 (d-f) and Supplementary Section IV), which aligned with the direction of unidirectional mode propagation around the ring. This alignment confirmed the topological nature of the mode, as the vorticity and mode propagation shared the same directional flow, affirming the system's robustness against defects.

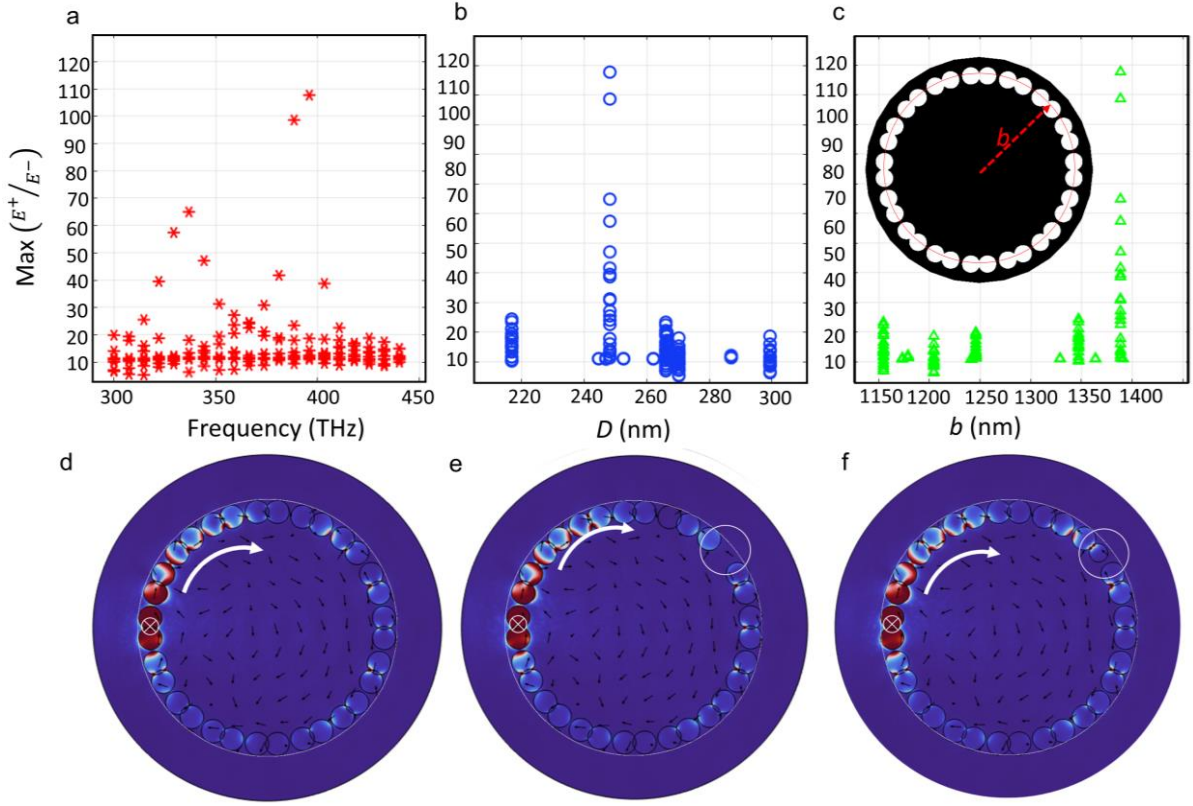

**Figure 5:** Symmetry-breaking excitation using local non-uniform excitation is employed to observe topological edge modes in a plasmonic chain with inherent symmetry. (a) The maximum electric field values for positive and negative, $|E^+|/|E^-|$, are determined based on key parameters, revealing optimal conditions for maximizing this ratio, occurring at: (a). Frequency of 380 THz, (b). At nanohole diameter $D$=248 nm. (c). The ring diameter $b$=1380 nm. (d-f) Display the spatial profile of the electric field's norm at a cross-sectional plane for the topological mode. The topological mode, protected by the ring geometry's symmetry and topology, remains robust despite local defects caused by (e) the absence of a nanoparticle or (f) changes in nanoparticle size. The white arrow likely indicates the total spin angular momentum at the cross-sectional plane, which represents the overall behavior of the system. The black arrows should show the local vorticity of spin angular momentum at specific points in the plane.

In conclusion, we have successfully demonstrated the power of physics-informed deep learning in the design of topological nanophotonic devices with targeted operational wavelengths. By embedding key physical principles in three steps into the learning framework, we significantly reduce computational overhead compared to traditional methods. This strategy enabled the swift identification of optimal device configurations, particularly those that exhibit robust one-way edge modes in perforated nanohole arrays. The introduction of non-planar wavefront excitations via electron beam enhances the functionality of these devices, offering novel means to selectively excite protected plasmonic modes that are otherwise inaccessible through conventional techniques. Our approach not only accelerates the discovery of novel topological devices but also sets a new benchmark for reducing design time and computational resources in nanophotonic systems. The potential applications of these innovations extend across fields, from optical communication to quantum technologies, where control over highly nonlinear systems is crucial. This work demonstrates how the integration of deep learning with domain-specific physics can transform photonic device engineering, unlocking scalable, high-performance designs for a variety of advanced technologies.

## Supporting Information

The Supporting Information is available free of charge at https://pubs.acs.org.

Detailed information on the interaction Hamiltonian in a two-level system, tight bonding Hamiltonian in SSH plasmonic chain, tight bonding Hamiltonian in SSH ring chain, visualization of the orbital angular momentum for both topologically nontrivial modes Tight bonding Hamiltonian in ring, and methods.

## Acknowledgement

This project has received funding from KiNSIS Early Career Award 2023.